\documentclass[review,3p]{elsarticle}

\usepackage{amsmath}
\usepackage{amssymb}
\usepackage{graphicx}
\usepackage{dcolumn}
\usepackage{bm}
\usepackage{stmaryrd}
\usepackage{wasysym}
\usepackage[figtopcap]{subfigure}
\usepackage{natbib}
\usepackage[numbers]{natbib}

\journal{Optics Communications}

\begin{document}

\begin{frontmatter}

\title{Atomic transitions of Rb, $D_{2}$ line in strong magnetic fields: hyperfine Paschen-Back regime}

\author{A. Sargsyan$^1$, A. Tonoyan$^{1,2}$, G. Hakhumyan$^1$, C. Leroy$^2$, Y. Pashayan-Leroy$^2$, D. Sarkisyan$^1$  }
\address{$^1$Institute for Physical Research, 0203, Ashtarak-2, Armenia}
\address{$^2$Laboratoire Interdisciplinaire Carnot de Bourgogne, UMR CNRS 6303, Universit\'e de Bourgogne, 21078 Dijon Cedex, France}

\begin{abstract}
An efficient $\lambda/2$-method ($\lambda$ is the resonant wavelength of laser radiation) based on nanometric-thickness cell filled with rubidium is implemented to study the splitting of hyperfine transitions of $^{85}$Rb and $^{87}$Rb $D_2$ lines in an external magnetic field in the range of $B =3$~kG -- 7~kG. It is experimentally demonstrated that at $B > 3$~kG from 38 (22) Zeeman transitions allowed at low $B$-field in
$^{85}$Rb ($^{87}$Rb) spectra in the case of $\sigma^+$ polarized laser radiation there remain only 12 (8)
which is caused by decoupling of the total electronic momentum $\textbf{J}$ and the nuclear
spin momentum $\textbf{I}$ (hyperfine Paschen-Back regime). Note that at $B > 4.5$~kG in the absorption spectrum these $20$ atomic transitions are regrouped in two completely separate groups of $10$ atomic transitions each. Their frequency positions and fixed (within each group) frequency slopes, as well as the probability characteristics are determined. A unique behavior of the atomic transitions of $^{85}$Rb  and
$^{87}$Rb labeled $19$ and $20$ (for low magnetic field they could be presented as transitions
$F_g=3, m_F=+3 \rightarrow F_e=4, m_F=+4$ and $F_g=2, m_F=+2 \rightarrow F_e=3, m_F=+3$, correspondingly) is stressed. The experiment agrees well with the theory. Comparison of the behavior of atomic transitions
for $D_2$ line compared with that of $D_1$ line is presented. Possible applications are described.
\end{abstract}

\date{\today}
\end{frontmatter}

\section{Introduction}
It is well known that in an external magnetic field $B$ the energy levels of atoms  undergo splitting
into a large number of Zeeman sublevels which are strongly frequency shifted, and simultaneously,
there are changes in the atomic transition probabilities~\cite{Tremblay,Alexandrov}.
Since ceasium and rubidium are widely used for investigation of optical and magneto-optical processes
in atomic vapors as well as for cooling of atoms, for Bose-Einstein condensation, and in a number of
other problems~\cite{Budker,Auzinsh2}, therefore, a detailed knowledge of the behavior of atomic levels
in external magnetic fields is of a high interest. The implementation of recently developed technique
based on narrowband laser diodes, strong permanent magnets and nanometric-thickness cell (NTC) makes
the study of the behavior of atomic transitions  in an external strong magnetic field simple and robust,
and allows one to study the behavior of any individual atomic transition of $^{85}$Rb and $^{87}$Rb atoms for $\textit{D}_1$ line~\cite{Sargsyan1,Sargsyan2}.\\
\indent Recently, a number of new applications  based on thin atomic vapor cells placed in a strong
magnetic field have been demonstrated: i) development of a frequency reference based on permanent magnets and
micro- and nano-cells
widely tunable over the range of several gigahertz by simple displacement of the
magnet; ii) optical magnetometers with micro- and/or  nano-metric spatial resolution ~\cite{Sargsyan1,Sargsyan2}; iii)
a light, compact optical isolator using an atomic Rb vapor in the hyperfine Paschen-Back regime
is presented in~\cite{Weller1,Weller2}; iv) it is demonstrated that the use of  Faraday rotation signal provides a simple
way to measure the atomic refractive index~\cite{Zentile}; v) widely tunable narrow optical resonances which are
convenient for a frequency locking of diode-laser radiation~\cite{Sarg2012}.\\
\indent Strong permanent magnets produce non-homogeneous magnetic fields. In spite
 of the strong inhomogeneity of the $B$-field (in our case it can reach $15$~mT/mm),
 the variation of \textit{B}
 inside the atomic vapor column is by several orders less than the applied $B$ value because of a
 small thickness of the cells. In case of micrometer thin cells with the thickness $L$ in the range
 of $10-50$~$\mu$m the spectral resolution is limited by the absorption Doppler line-width of an individual
 atomic transition (hundreds of megahertz). If the frequency distances between Zeeman
 sublevels are small a big number of atomic transitions are strongly overlapped
 and it makes absorption spectra very complicated. Fortunately, as demonstrated for Cs $D_{2}$ line, at
 strong ($B$ $>4$~kG) magnetic fields  $16$ atomic transitions in the absorption spectrum (there are $54$
 atomic transitions in moderate magnetic fields for circular polarization of the excitation field)
  are frequency separated from each other by a value slightly larger than the absorption Doppler
  line-width of an individual atomic transition~\cite{Sarg2013}. That's why in this case even the use of micrometer
  thin cells allows one to separate practically all $16$ atomic transitions (so called hyperfine
  Paschen-Back regime (HPB)).\\
\indent Note that even for such large values as $B>4$~kG, the atomic transitions of $^{87}$Rb and
$^{85}$Rb $D_{2}$ lines are
strongly overlapped, so pure isotope $^{87}$Rb and $1$~mm-atomic vapor cell have been used to separate eight
Zeeman transitions~\cite{Weller1,Weller2,Zentile}. However with this technique even in the case of using pure isotope $^{85}$Rb, the atomic lines will be  strongly overlapped.\\
\indent Although, the HPB regime was discovered many decades earlier (see Refs. in~\cite{Alexandrov,Umfer,Olsen}),
however the implementation of recently developed setup based on narrowband laser diodes, strong
permanent magnets and NTC makes these studies simple and robust, and allows one to study the behavior
of any individual atomic transition of $^{85}$Rb and $^{87}$Rb atoms; the simplicity of the system also makes
it possible to use it for a number of applications.\\
\indent In this paper we present (for the first time to our best knowledge), the results of experimental
and theoretical studies of the Rb $D_{2}$
 line transitions (both $^{87}$Rb and $^{85}$Rb
 are presented) in a wide range
of magnetic fields, namely for $3$~kG~$<  B  <$~$7$~kG. It is experimentally demonstrated that in the case of $B>4.5$~kG and $\sigma^{+}$
polarized laser radiation, there remain only $20$ Zeeman transitions. In the absorption spectrum these $20$ transitions are regrouped in  two separate groups each of $10$
atomic transitions (HPB regime), while there are $60$ allowed Zeeman
transitions at low $B$-field.

\section{EXPERIMENTAL DETAILS}
\subsection{Nanometric-thin cell construction}
\indent The design of a nanometric-thin cell  NTC is similar to that of extremely thin cell
described earlier~\cite{Sarkisyan2001}. The modification implemented in the present work is as follows.
The rectangular $20$~mm $\times$ $30$~mm,~$2.5$~mm-thick window wafers polished to $<1$~nm surface roughness are
fabricated from commercial sapphire (Al$_{2}$O$_{3}$), which is chemically resistant
to hot vapors (up to $1000\,^{\circ}$C) of alkali metals.
The wafers are cut across the $c$-axis to
minimize the birefringence. In order to exploit variable vapor column thickness, the cell
is vertically wedged by placing a $1.5$~$\mu$m-thick platinum spacer strip between the windows
at the bottom side prior to gluing. The NTC is filled with a natural rubidium
$(72.2\, \%$ $^{85}$Rb and $27.8\,\% $ $^{87}$Rb). A thermocouple is attached to the sapphire side
arm at the boundary of metallic Rb to measure the temperature, which determines the vapor pressure.
The side arm temperature in present experiment was $120\, ^{\circ}$C, while the windows
temperature was kept some $20\, ^{\circ}$C higher to prevent condensation. This temperature regime
corresponds to the Rb atomic number density $N =2\cdot10^{13}$~cm$^{-3}$. The NTC operated with
a special oven with two optical outlets. The oven (with the NTC fixed inside) was rigidly attached
to a translation stage for smooth vertical movement to adjust the needed vapor column thickness
without variation of thermal conditions. Note, that all experimental results have been
obtained with Rb vapor column thickness $L=\lambda/2=390$~nm. For more details see~\cite{Keaveney}.\\

\subsection{Experimental setup}
\begin{figure}[hbtp]
\centering
\includegraphics[scale=0.5] {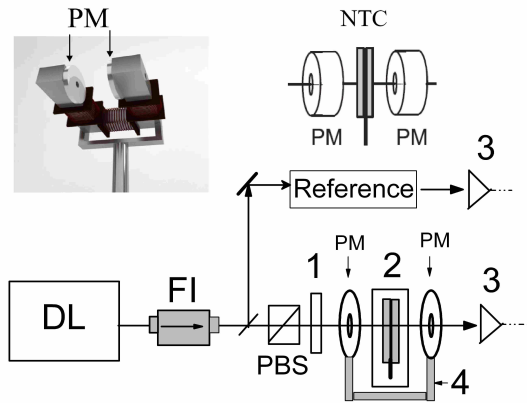}
\caption{
Sketch of the experimental setup. DL - tunable diode laser, FI - Faraday isolator, $1$ - $\lambda/4$
plate, PBS - Polarizing Beam Splitter, $2$ - NTC in the oven, Reference -(FR)- auxiliary
Rb NTC providing $B = 0$ reference spectrum, PM - permanent  magnets, $3$ - photo-detectors,
$4$ - metallic magnetic core.
  }
\label{fig:NTC}
\end{figure}

\indent Figure~\ref{fig:NTC} presents the experimental scheme for the detection of the absorption spectrum of
the nano-cell filled with Rb. It is important to note that the implemented  $\lambda/2$-method" exploits
strong narrowing of absorption spectrum at $L = \lambda/2$ as compared with the case of an ordinary
cm-size cell~\cite{Sarg2013,Sarkisyan2001}. Particularly, the absorption line-width for Rb $D_2$ line ($L = \lambda/2= 390$~nm)
 reduces to about of $200$~MHz (FWHM), as opposed to that in an ordinary cell (about of $500$~MHz).
In the experiment we used the radiation of a continuous
wave narrowband diode laser with the wavelength of $780$~nm and the width of $10$~MHz.
In the current experiment we had the choice to use  a diode laser with the line-width of $\approx1$~MHz, however the
mode hope free region is only of $5$~GHz, which is too small to register all $20$ atomic transitions. Meanwhile
with the laser used in the experiment the mode hope free tuning range is $40$~GHz. The linearity of the scanned frequency was tested
by simultaneously recorded transmission spectra of a Fabry-P\'{e}rot etalon (not shown). The nonlinearity
has been evaluated to be about $1\%$ throughout the spectral range.\\
\indent The strong magnetic field was produced by two $\varnothing$ $50$~mm permanent magnets (PM) with $3$~mm
holes (to allow the radiation to pass) placed on the opposite sides of the NTC oven and separated by a
distance that was varied between $40$ and $25$~mm (see the upper inset in Fig.~\ref{fig:NTC}). The magnetic field
was measured by a calibrated Hall gauge. To control the magnetic field value, one of the magnets
was mounted on a micrometric translation stage for longitudinal displacement. In the case where the
minimum separation distance is of $25$~mm, the magnetic field $B$ produced inside the NTC reaches $3600$~G.
To enhance the magnetic field up to $6$~kG, the two PM were fixed to a metallic magnetic core with a cross
section of $40$~mm x $50$~mm. Additional form-wounded Cu coils allow for the application of extra $B$-fields
(up to $\pm \,1$~kG) (see the inset of Fig.~\ref{fig:NTC}).\\

\indent The beam with $\sigma^{+}$ circular polarization  was formed by a $\lambda/4$ plate.
The beam was focused by a lens ($F = 20$~cm) on the NTC to create a spot size ($1/e^2$ diameter,
i.e. the distance where the power drops to $13.5$~\% of its peak value) in the cell of $d =0.6$~mm
and then collimated by a second lens (not shown in Fig.~\ref{fig:NTC}).
To form
the frequency reference (from which the frequency shifts were measured), a part of the laser beam was
directed to a unit composed of an additional NTC with $L = \lambda/2$. The absorption spectrum
of the latter at the atomic transition $F_g = 1 \rightarrow F_e = 1, 2$ served as a reference (another
weak transition $F_g = 1 \rightarrow F_e = 0$ is not well seen)~\cite{Sarg2012}.

\section{EXPERIMENTAL RESULTS and DISCUSSIONS}
\subsection{Magnetic field $B < 3$~kG}
\indent In case of relatively low magnetic fields ($\sim 1$~kG) there are $60$ allowed Zeeman
transitions  when circular laser radiation excitation is used,
with $22$ atomic transitions belonging to $^{87}$Rb, and $38$ transitions belonging
to $^{85}$Rb $D_{2}$ line. These numerous atomic transitions are strongly
overlapped and can be partially resolved in case of using  $^{87}$Rb or $^{85}$Rb isotope. When using
 natural Rb, the implementation of $\lambda/2$-method allows one to
resolve practically any individual atomic transition only for $B \geq 3$~kG.\\
\indent The reduction of the total number of allowed atomic transitions at high magnetic
fields down to $20$  is caused by the effect of strong reduction of the atomic transitions
probabilities for $40$ transitions (such type of calculations for Rb $D_1$ line are
presented in~\cite{SargPhy}). Note, that for $B \gg B_0$ the number of allowed transitions can be simply
obtained from the diagrams shown in Fig.~\ref{fig:Fig4}.\\
\begin{figure}[hbtp]
\centering
\includegraphics[scale=0.3] {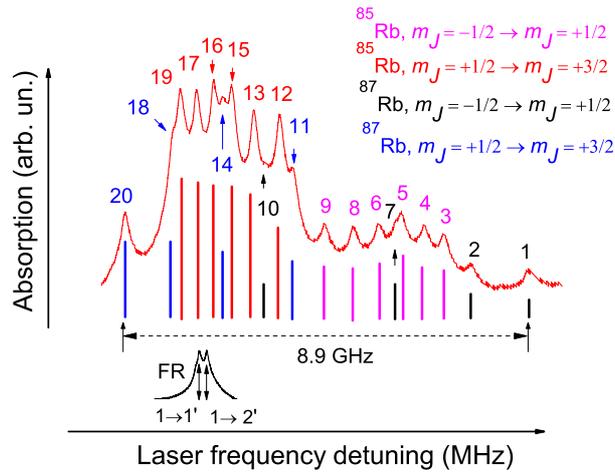}
\caption{
Absorption spectrum of Rb NTC with $L =\lambda/2$ for $B = 3550$~G and $\sigma^{+}$ laser excitation.
 The bottom curve (FR) is the absorption spectrum of the reference NTC showing the positions of $^{87}$Rb $1 \rightarrow 1^{\prime}, 2^{\prime}$
 transitions for $B = 0$ (the frequency separation is $157$~MHz). In the upper corner the corresponding
 atomic transitions are indicated. The absolute
value of the peak absorption of the transition labeled as $1$ is $\sim 0.3$\%.  }
\label{fig:fig2}
\end{figure}
\indent In Fig.~\ref{fig:fig2} the absorption spectrum of Rb NTC with $L= \lambda/2$ for $B = 3550$~G and $\sigma^{+}$ laser
excitation is shown. The laser power is $10$~$\mu$W. For Rb $D_{2}$ line there are $20$ atomic absorption resonances located at the atomic transitions. Among these transitions $12$ belong to $^{85}$Rb, and $8$
transitions belong to $^{87}$Rb.\\
\indent The atomic transition pairs labeled ($19,18$) and ($7,5$) are
strongly overlapped (although in the case of strongly expanded spectrum the peaks belonging to
the corresponding transitions are well detected), while the other $16$ transitions are overlapped
partially, and the positions of the absorption peaks are well seen. Thus, the fitting of the
absorption spectrum with $20$ atomic transitions  is not a difficult problem. The vertical bars
presented in Fig.~\ref{fig:fig2}  indicate the frequency positions and the magnitudes for individual transitions
between the Zeeman sublevels as given by numerical simulations using the model described below.
The corresponding atomic transitions presented by the vertical bars are indicated in the upper corner of Fig.~\ref{fig:fig2}.
\subsection{Magnetic field $B > 3$~kG : hyperfine Paschen-Back (HPB) regime}
\indent In case of strong ($B > 3$~kG) magnetic fields, the frequency separation
between the atomic transitions increases, which allows one to separate practically
any individual transition
by using $\lambdaλ/2$-method. A remarkable value of the magnetic field is $4.5$~kG, since
at $B > 4.5$~kG, $20$ atomic transitions are regrouped to form two separate groups
of ten transitions each and the frequency interval between these two groups increases with the magnetic field (see below).\\
 \begin{figure}[t!]
 \subfigure[]{
 \resizebox{0.45\textwidth}{!}{\includegraphics{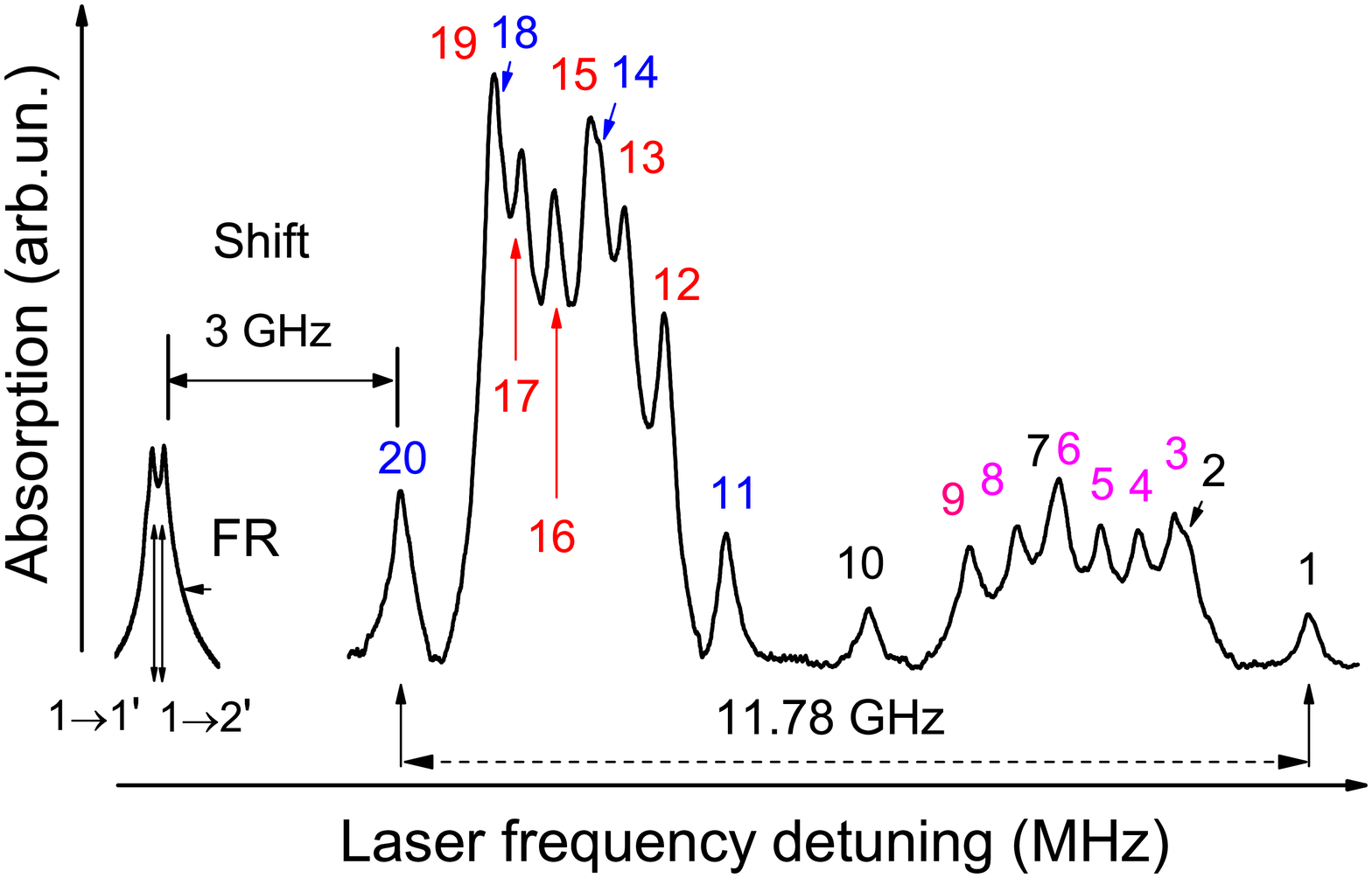}
 }
 \label{fig:Fig3a}
 } \vspace{0.005cm}
 \subfigure[]{
 \resizebox{0.45\textwidth}{!}{\includegraphics{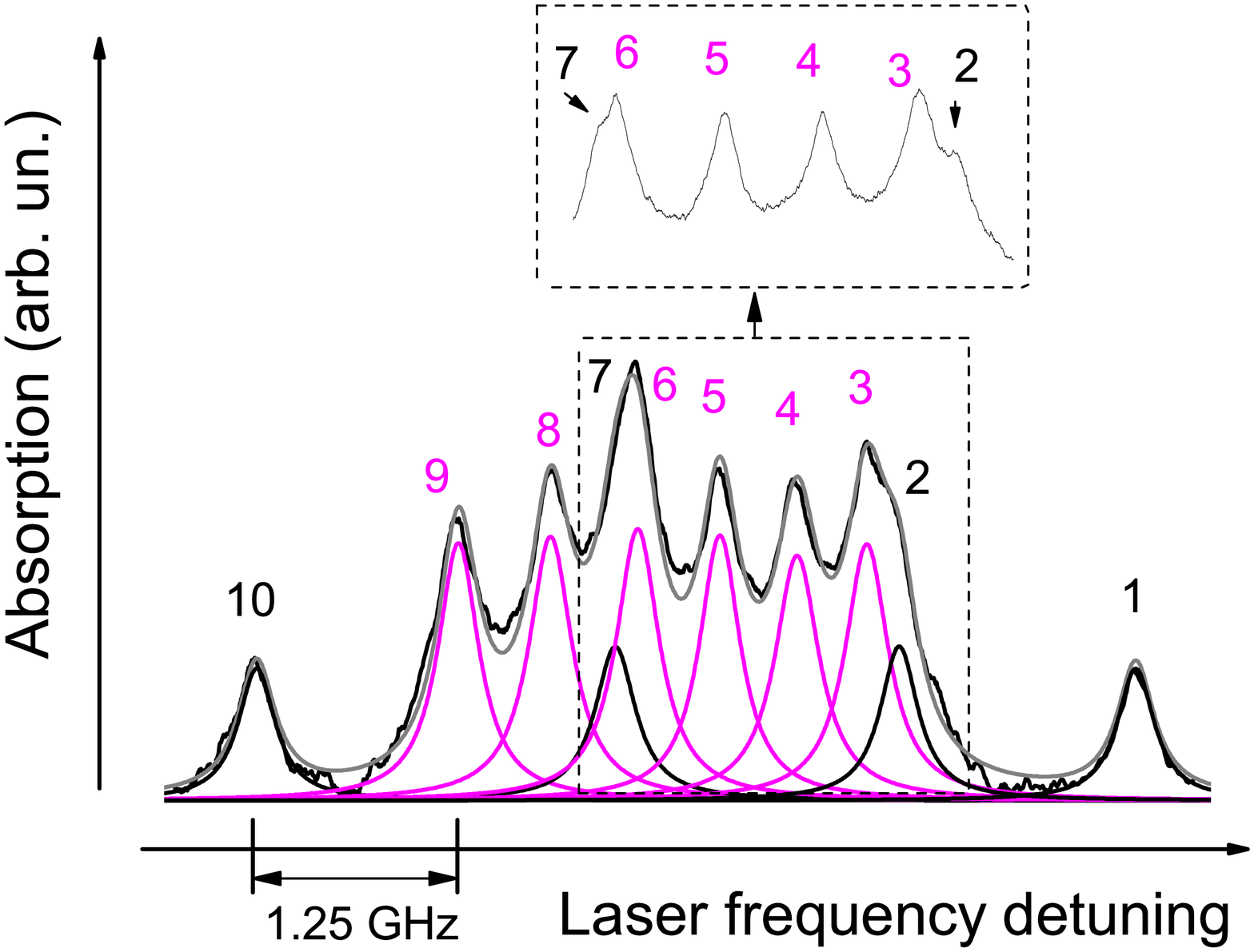}
 }
 \label{fig:Fig3b}
 } \vspace{0.005cm}
  \subfigure[]{
 \resizebox{0.45\textwidth}{!}{\includegraphics{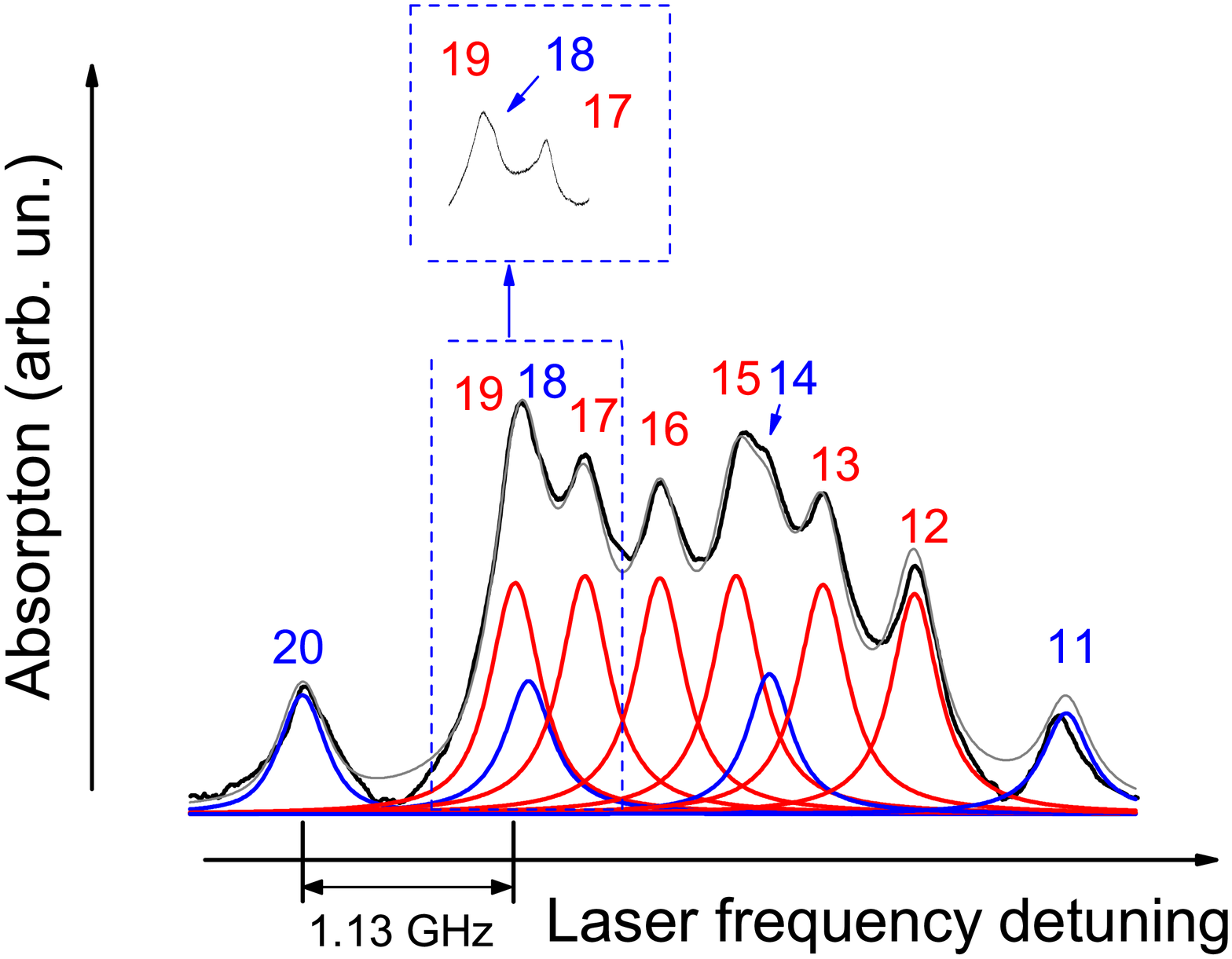}
 }
 \label{fig:Fig3c}
 }
  \caption{a) Absorption spectrum of Rb NTC with $L =\lambda/2$ for $B = 6850$~G and $\sigma^{+}$ laser excitation.
 For the  transition labels, see Fig.~\ref{fig:Fig4}. The left curve is the absorption spectrum of the reference
 NTC showing the positions of $^{87}$Rb $1 \rightarrow1^{\prime}$, $2^{\prime}$ transitions for $B = 0$ (the frequency
 separation is $157$~MHz) and the frequency shift of the atomic transition labeled $20$ with respect to $1 \rightarrow2^{\prime}$.
 b) The fragment of the absorption spectrum presented in Fig.~\ref{fig:Fig3}\subref{fig:Fig3a}. The group contains the atomic
transitions labeled $1-10$ which are fitted with the pseudo-Voigt profiles, with the line-width
(Full Width Half Maximum)  of $250$~MHz;  the inset shows an expanded view of the part of the
experimental results limited by the dashed rectangle. c) The fragment of the absorption spectrum
presented in Fig.~\ref{fig:Fig3}\subref{fig:Fig3a} containing the atomic transitions labeled $11-20$ which are fitted with the
pseudo-Voigt profiles.  }
 \label{fig:Fig3}
 \end{figure}
\indent In Fig.~\ref{fig:Fig3} the absorption spectrum of Rb NTC of $L= \lambda/2$ for $B = 6850$~G and $\sigma^{+}$ laser
excitation is shown. There are still $20$ atomic absorption resonances of
Rb $D_{2}$ line located at the atomic transitions. Among these transitions, $12$ belong to $^{85}$Rb,
and $8$ transitions belong to $^{87}$Rb. Atomic transition pairs labeled ($19,18$), ($15,14$)
and ($3,2$) are overlapped (although in the case of strongly expanded spectrum the peaks
belonging to the corresponding transitions are well detected, see Fig.~\ref{fig:Fig3}\subref{fig:Fig3c})
while the other $14$
transitions are overlapped partially, and the positions of the absorption peaks of the individual
transitions are well detected. The left curve presents the absorption spectrum of the reference NTC
with $L =\lambda/2$ showing the positions of the $^{87}$Rb, $F_{g} =1 \rightarrow F_{e} = 1, 2$ transitions for $B$ $= 0$ (the
frequency shift of the transitions is determined with respect to $1 \rightarrow 2'$ transition).\\
\indent Figure~\ref{fig:Fig3}\subref{fig:Fig3b} shows the fragment of the spectrum (presented in Fig.~\ref{fig:Fig3}\subref{fig:Fig3a}) for the
atomic transitions labeled $1 \rightarrow 10$, where the transitions labeled $1, 2, 7$ and $10$ belong to $^{87}$Rb,
while the transitions labeled $3 -6, 8$, and~$9$ belong to $^{85}$Rb. The fitting (with the
pseudo-Voigt profiles ~\cite{Sarg2013}) is justified through the following advantageous property of the
$\lambda/2$-method: in case of a weak absorption, the absorption coefficient $A$ of an individual
transition component is proportional to $\sigma NL$, where $\sigma$ is the absorption cross-section
proportional to $d^{2}$ (with $d$ being the matrix element of the dipole moment), $N$ is the atomic density,
and $L$ is the thickness.  Measuring the ratio of $A_{i}$ values for different individual transitions,
it is straightforward to estimate their relative probabilities (line intensities).\\
\indent The fragment of the spectrum (presented in Fig.~\ref{fig:Fig3}\subref{fig:Fig3a}) is shown in Fig.~\ref{fig:Fig3}\subref{fig:Fig3c} for the atomic
transitions labeled $11-20$, where the transitions labeled $11,\, 14,\, 18$ and $20$  belong to $^{87}$Rb,
while the transitions labeled $12,\, 13,\, 15,\, 16,\, 17$, and $19$  belong to $^{85}$Rb.\\
\indent It is important to note that, as seen from Fig.~\ref{fig:Fig3}, the absorption peak numbered $1$
is the most convenient for magnetic field measurements, since it is not overlapped with any
other transition in the range of  $1-10$~kG   (see also Fig.~\ref{fig:Fig4}), while having a strong detuning
value  in the range of $2-2.3$~MHz/G.\\
\subsection{The manifestation of HPB regime.}
\begin{figure}[t!]
 \subfigure[]{
 \resizebox{0.45\textwidth}{!}{\includegraphics{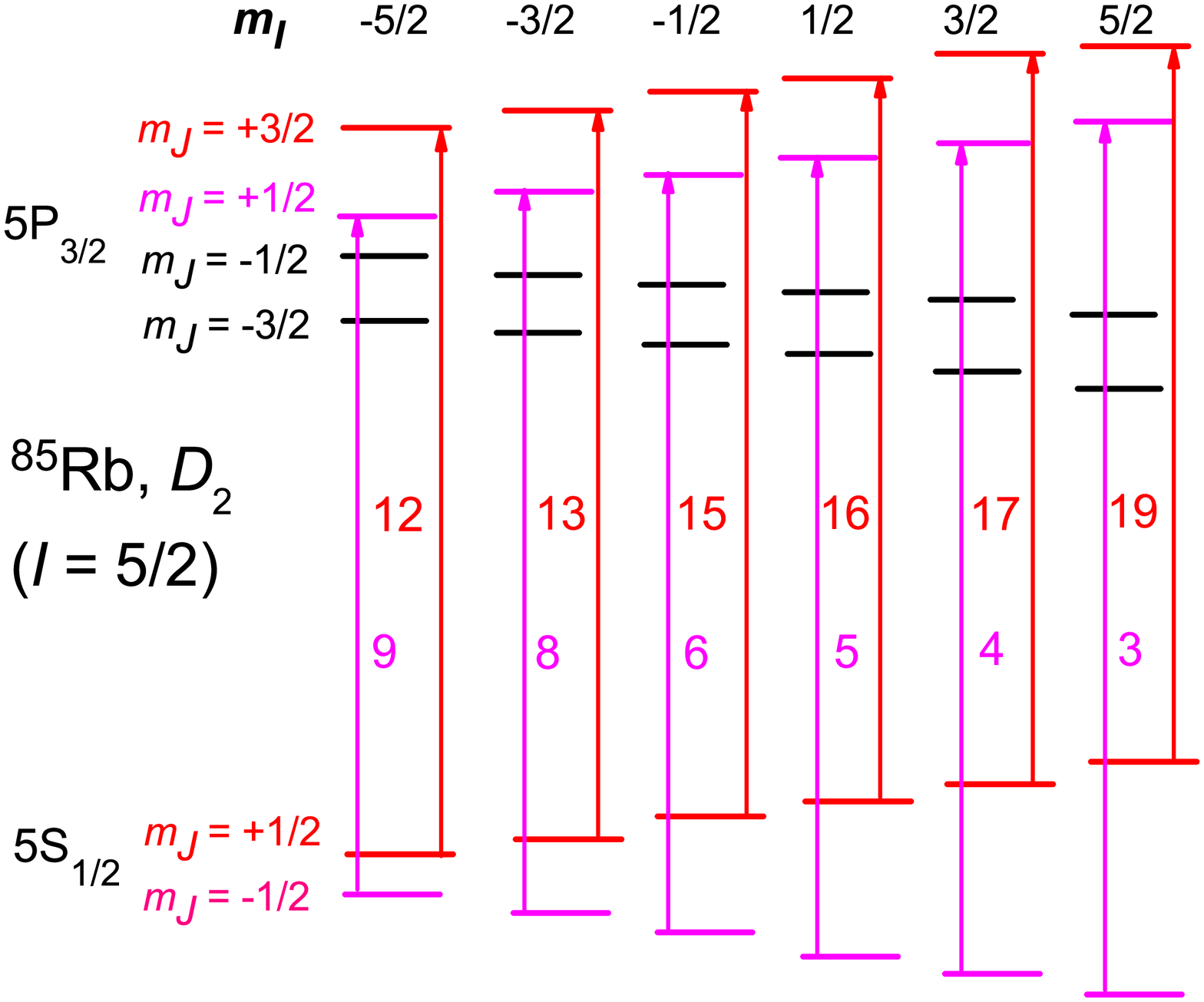}
 }
 \label{fig:Fig4a}
 } \hspace{0.005cm}
 \subfigure[]{
 \resizebox{0.45\textwidth}{!}{\includegraphics{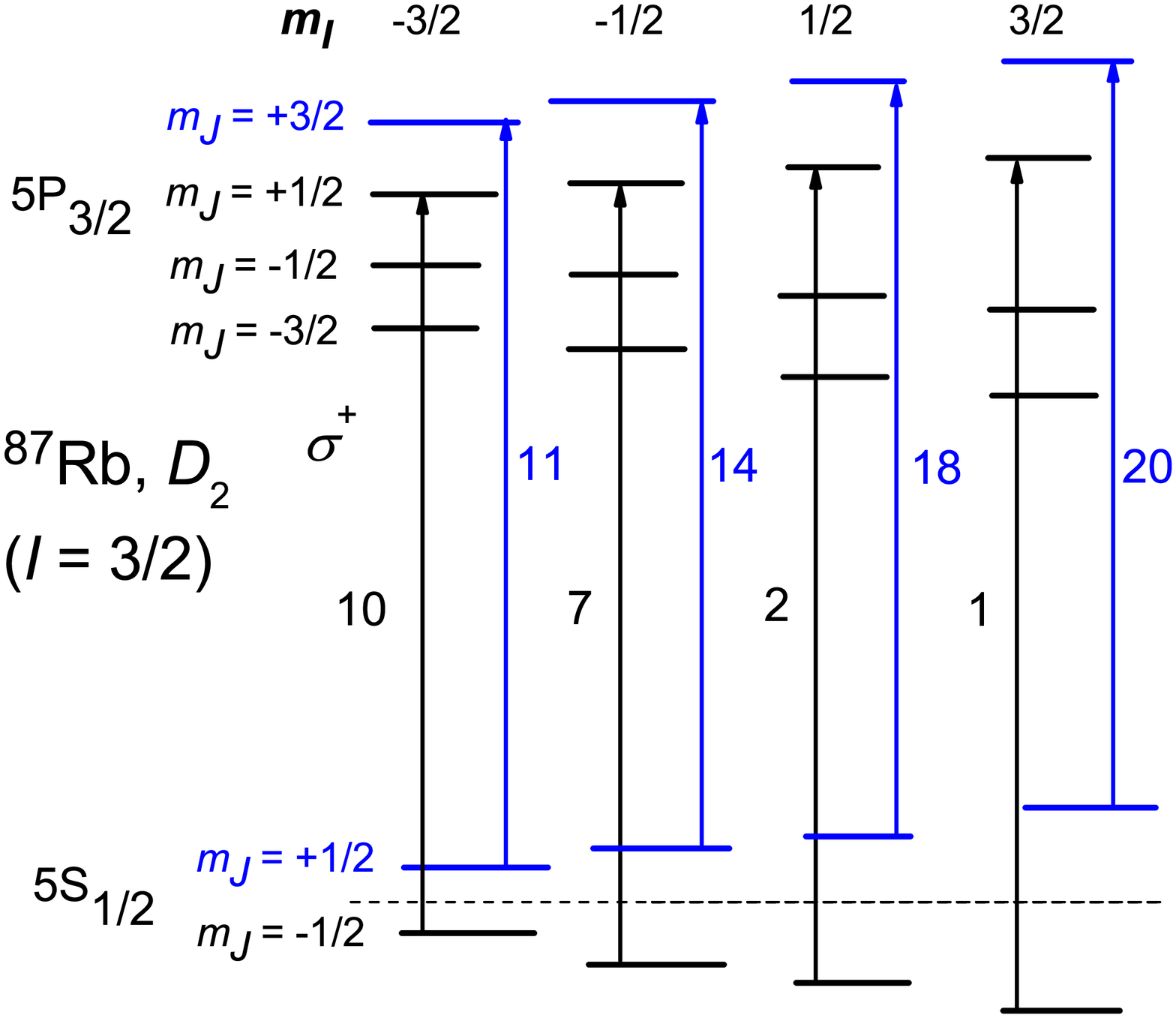}
 }
 \label{fig:Fig4b}
 }
 \caption{ a) Diagram of $^{85}$Rb, $D_{2}$ line ($I = 5/2$) transitions for $\sigma^{+}$ laser excitation in HPB regime.
 The selection rules: $\Delta _{m_J} = +1; \Delta _{m_I} = 0$. Therefore, there are $12$ atomic transitions marked by the
 respective numbers $3-6,8,9,12,13,15,16,17$, and $19$. b) Diagram of $^{87}$Rb, $D_{2}$ line ($I = 3/2$) transitions
 for $\sigma^{+}$ laser excitation in HPB regime. Due to the selection rules there are $8$ atomic
 transitions marked by the respective numbers $1, 2, 7, 10, 11, 14, 18$, and $20$.
  }
 \label{fig:Fig4}
 \end{figure}

\indent The magnetic field required to decouple the electronic total angular momentum $\textbf{\textit{J}}$ and the
nuclear magnetic momentum $\textbf{\textit{I}}$ is given by $B \gg B_{0} = A_{hfs}/\mu_{B}$.  For $^{85}$Rb and   $^{87}$Rb it is estimated to be approximately
equal to $B_{0}$ $(^{87}$Rb) $\approx$ 2~kG, and  $B_{0}$ $(^{85}$Rb) $\approx$ $0.7$~kG, where $A_{hfs}$ is the ground-state hyperfine coupling
coefficient for $^{87}$Rb and $^{85}$Rb and $\mu _{B}$ is the Bohr magneton~\cite{Alexandrov, Var}.\\
\indent For such strong magnetic fields when $\textbf{\textit{I}}$ and $\textbf{\textit{J}}$ are decoupled (HPB regime),  the eigenstates of the
Hamiltonian are described in the uncoupled basis of $J$ and $I$ projections $(m_{J}; m_{I})$.  Fig.~\ref{fig:Fig4}\subref{fig:Fig4a} shows
$12$ atomic transitions of $^{85}$Rb labeled  $3 - 6, 8, 9, 12, 13,15, 16, 17$ and $19$ for the
case of $\sigma^{+}$ polarized laser excitation in the HPB regime and $8$  transitions of $^{87}$Rb labeled
$1, 2, 7, 10, 11,14,18$ and $20$.\\
 \begin{figure}[hbtp]
\centering
\includegraphics[scale=0.3] {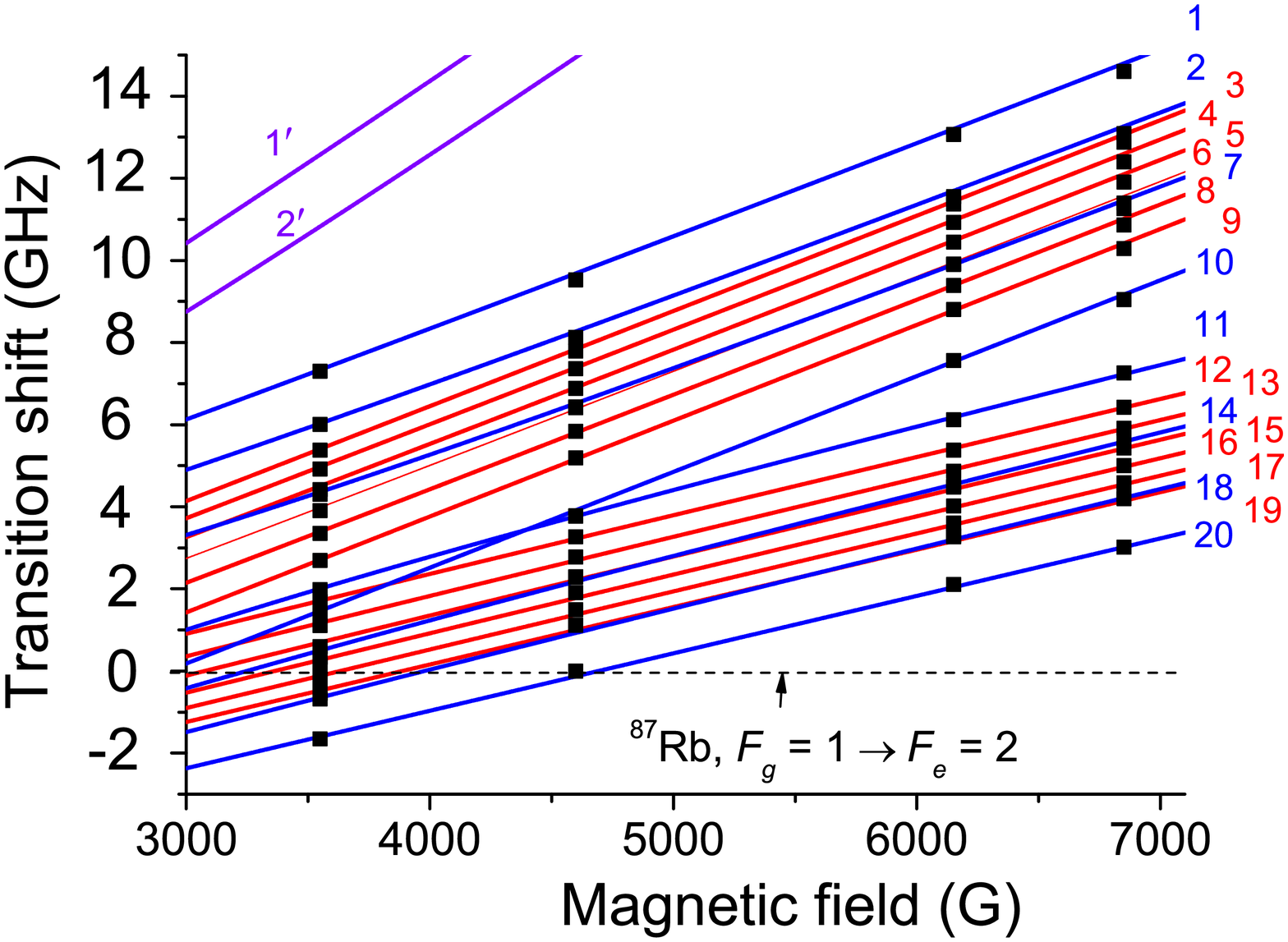}
\caption{
 Frequency positions of the Rb, $D_{2}$ line atomic transitions $1 - 20$ versus the magnetic
 field. Solid lines are the calculated curves and black squares are the experimental
 results (with an error of $3$\%). At $B > 4.5$~kG, the transitions are regrouped to form
 two groups of ten transitions each. For $B\gg B_{0}$ the frequency slope of the $1$-st group
 ($1-10$ transitions) is $s_{1} \approx  2.33$~MHz/G, and for the second group ($11-20$ transitions)
$s_{2} \approx 1.39$~MHz/G. Two upper curves show $1^\prime$ and $2^\prime$  belonging to the $^{87}$Rb,
$D_2$, $F_g=1 \rightarrow F_e=3$ transitions, with the probability reducing to zero for $B > 6$~kG.
Note, that magnetic field-induced strong modification of the probabilities for atomic transitions
(which are forbidden by selection rules at $B=0$) has been demonstrated in~\cite{Sarg2011,Papoyan}.
  }
\label{fig:fig5}
\end{figure}
\indent Simulations of magnetic sublevel energy and relative transition probabilities for Rb $D_{2}$ line are well
known, and are based on the calculation of dependence of the eigenvalues and eigenvectors of the Hamiltonian
matrix on magnetic field for the full hyperfine structure manifold~\cite{Alexandrov,Auzinsh2,Sargsyan1,Sark2001,Sarg2011,SargPhy}. The calculated dependence
of atomic transition probabilities (shown in Fig.~\ref{fig:fig6}) and frequency shifts vs $B$ (shown in Fig.~\ref{fig:fig5}) are obtained
using formulas ($1$)-($7$) from work [$11$] and are omitted in the paper due to their bulkiness.  Particularly, recently it has been demonstrated that the calculations using the above mentioned formulas perfectly well describe the experimental observation of  a giant modification of the atomic probabilities of the Cs $D_2$ line $F_g=3 \rightarrow F_e=5$ transitions (which are forbidden for $B=0$) in strong magnetic fields~\cite{Papoyan}. \\
\indent It is important to note that for $B\gg B_{0}$  the energy of the ground $5S_{1/2}$ and upper $5P_{3/2}$ levels
for Rb $D_{2}$ line is given by the following  equation~\cite{Var}:
\begin{equation}
\begin{array}{r}
E_{|Jm_{J} Im_{I} \rangle}= A_{hfs} m_J m_I +  B_{hfs} \frac{3(m_J m_I)^2 +
 \frac{3}{2} m_J m_I - I(I + 1) J (J + 1)}{2J (2J - 1) (2I - 1)} + \mu_B (g_J m_J + g_I m_I) B_z.
\end{array}
\label{eq:E}
\end{equation}
\indent The values for nuclear $(g_{I})$ and fine structure $(g_{J})$ Land\'{e} factors and hyperfine constants $A_{hfs}$ and $B_{hfs}$
are given in ~\cite{Var}. Note, that Eq.~\eqref{eq:E} gives correct frequency positions of the components $1-20$ with an
inaccuracy of $2\%$ practically when $B \geq 10 B_{0}$, i.e. $B \geq 6-7$~kG for $^{85}$Rb and $B \geq 20$~kG for $^{87}$Rb~\cite{SargPhy}.\\
\indent Fig.~\ref{fig:fig5} illustrates the frequency positions (i.e. frequency shifts) of the components $1 - 20$ as functions
of the magnetic field $B$. The theoretical curves are shown by solid lines. The black squares are the
experimental results  which are in a good agreement with the theoretical curves (with an error of $3$\%). As seen,
the atomic transitions are regrouped at $B > 4500$~G to
form two new sets of ten transitions each. Note, that the frequency separation between these two groups
increases with $B$. The dashed line denotes the frequency position of the $^{87}$Rb, $F_{g}=1\rightarrow F_{e}=2$ transition for $B=0$.\\
\begin{figure}[hbtp]
\centering
\includegraphics[scale=0.3] {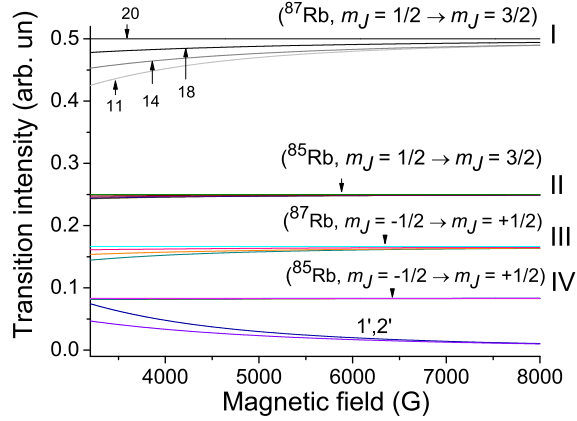}
\caption {
Intensity (probability) of the atomic transitions: ($I$-st group) $^{87}$Rb, $\textit{D}_2$ line transitions
labeled $11,14,18$, and $20$; ($II$-nd group) $^{85}$Rb, $\textit{D}_2$ line transitions
labeled $12,13,15-17$, and $19$; ($III$-rd group) $^{87}$Rb, $\textit{D}_2$ line transitions
labeled $1, 2, 7$, and $10$;  ($IV$-th group) $^{85}$Rb, $\textit{D}_2$ line
transitions labeled $3, 4, 5, 6, 8$ and $9$. The transition probabilities differ significantly at
low $\textbf{\textit{B}}$-fields but tend to the same value within the group at $B \gg B_{0}$.
Two lower curves show $1^\prime$ and $2^\prime$  belonging to the $^{87}$Rb, $\textit{D}_2$, $F_{g}=1 - F_{e}=3$
transitions, with the probability reducing to zero for $B > 6$~kG.
  }
\label{fig:fig6}
\end{figure}
\indent The experimental values of the slopes (the frequency shift with respect to the magnetic
field) at $B = 7$~kG are $s_{1}\approx 2.29$~MHz/G  (for $^{85}$Rb transitions this value is slightly larger,
while for $^{87}$Rb this value is slightly smaller) and $s_{2}\approx 1.42$~MHz/G  (for $^{85}$Rb transitions this value
is slightly smaller, while for $^{87}$Rb this value is slightly larger) for the groups $1$-$10$ and $11$-$20$,
respectively. It is noteworthy that the slope value for $^{87}$Rb and $^{85}$Rb (inside the same group, see Fig.~\ref{fig:Fig4})
are nearly equal to each other when the initial and final energy levels are the same (see below).\\
\indent The slopes of the transitions for the fields $B\gg B_{0}$ can be easily found from Eq.~\eqref{eq:E}
as~$s_{1} =$ $ [   g_{J} (P_{3/2}) m_{J} -$ $  g_{J} (S_{1/2}) m_{J} ] $~$ \mu _{B} / B\approx 2.33$~MHz/G and
$s_{2} = \left[   g_{J} (P_{3/2})   m_{J} -  g_{J} (S_{1/2}) m_{J} \right ] \mu _{B} / B$ $ \approx 1.39$~MHz/G
for the groups $1-10$ and $11-20$, respectively (as $g_{I} \ll g_{J}$, we ignore $g_{I} m_{I}$ contribution).
Consequently, at $B \geq 20$~kG (when the condition of HPB is fully satisfied for $^{87}$Rb atoms, too), the slope for
the group  $1-10$  increases slightly (to $s_{1}$), while the slope for the group  $11-20$
decreases slightly to  $s_{2}$. In addition, one can easily find from Eq.~\eqref{eq:E} the frequency
intervals between the components within each group.\\
\indent Figure~\ref{fig:fig6}  presents the theoretical values of  ``$1 - 20$'' atomic transitions probabilities (intensities)
 in the fields of $5$-$7$~kG. Let us compare the experimental results presented in Fig.~\ref{fig:Fig3} obtained
 for $B=6850$~G with the theoretical calculations of the atomic transitions probabilities shown in Fig.~\ref{fig:fig6}.
 The atomic transitions labeled 3, 4, 5, 6, 8 and 9 of $^{85}$Rb shown in Fig.~\ref{fig:Fig3}\subref{fig:Fig3b} have the same amplitudes (probabilities) with inaccuracy less than $5$\% and this is in a good agreement with the theoretical curves shown in the IV-th group in Fig.~\ref{fig:fig6}. The atomic transitions labeled 12, 13, 15 -- 17, and 19 shown in Fig.~\ref{fig:Fig3}\subref{fig:Fig3c} have the same amplitudes (probabilities) with inaccuracy less than 5\% and this is in a good agreement with the theoretical curves shown in the II-nd group in Fig.~\ref{fig:fig6}. It is easy to see that there is a  similar good agreement between the
 amplitudes (probabilities) for the atomic transitions of $^{87}$Rb shown in Fig.~\ref{fig:Fig3}\subref{fig:Fig3b} and \subref{fig:Fig3c} with the theoretical curves shown in the I-st and III-rd groups presented in Fig.~\ref{fig:fig6}. Note, that the probabilities of the transitions for  $^{87}$Rb inside the same group $1$-$10$ or $11$-$20$
are nearly two times larger than the probabilities for $^{85}$Rb inside the same group. However, since
for natural rubidium the atomic density ratio  $N(^{85}$Rb)$/ N(^{87}$Rb)$\, \approx 2.6$,  the peak absorption
of the atomic transitions for  $^{85}$Rb is nearly $1.5$ times larger than that for $^{87}$Rb (Fig. 6).\\
\indent It is worth to note that the probability of the atomic transition for  $^{87}$Rb labeled $20$
(for low magnetic field it could be presented as transition $F_g=2, m_F = +2 \rightarrow F_e = 3, m_F = +3$)  is the same in the whole range of magnetic field from zero up to $10$~kG. This is caused by the absence of Zeeman sublevels
with $F_g=1, m_F =+2$ and $F_e=2, m_F=+3$, since the perturbation induced by the magnetic field couples only sublevels with $\Delta m_F = 0$ which satisfies the selection rules $\Delta L=0$, $\Delta J=0$, $\Delta F=±1$, where $L$ is the orbital angular momentum and $F$ is the total atomic angular momentum (see formula (2) from~\cite{Sark2001}). Thus, probability modification is possible only for transitions between sublevels each of which is coupled with another transition according to the selection rules presented above. Due to the similar reason the probability of the atomic transition for $^{85}$Rb labeled $19$ (it could be presented as transition $F_g=3, m_F=+3 \rightarrow F_e=4, m_F=+4$) is also the same in the range of magnetic fields from zero up to $10$~kG. Since for the transitions labeled $19$
and $20$ the absolute value of the probability could be calculated from~\cite{Var}, thus using the experimental results presented in Fig.~\ref{fig:Fig3} the absolute value of the probabilities for the other atomic transitions (modified by magnetic field) can be calculated as well. Also, due to the above mentioned reason the frequency shifts of transitions labeled $19$ and $20$ as a function of magnetic field is simply linear with a fixed slope of $s =1.39$~MHz/G.
Note, that the reduction of the total number of Rb, $D_{2}$  transitions to strictly $20$ in
strong magnetic fields (which are well described by the diagrams presented in Fig.~\ref{fig:Fig4}), as well as
the behavior of the slopes $s_{1}$
and $s_{2}$   of $^{85}$Rb and  $^{87}$Rb (which are close to the values obtained by Eq.~\ref{eq:E}) is the
manifestation of the hyperfine Paschen-Back regime.\\
\indent Note, that there are three main distinctions in the behavior of atomic transitions  for $D_2$ line as compared with
the behavior for $D_1$ line~\cite{Sargsyan1,Sargsyan2}.\\
i)	At $B > 4.5$~kG in the absorption spectrum $20$ well resolved atomic transitions are regrouped in two completely separate groups of $10$ atomic transitions each and frequency separation between two groups increases with magnetic field. Meanwhile, for $D_1$ line there are only $10$ atomic transitions forming one group.\\
ii)	There are two remarkable atomic transitions for $D_2$ line: for $^{87}$Rb atom, the transition
labeled $20$ and for $^{85}$Rb atom, the transition labeled $19$. In a wide region of magnetic
fields from zero up to $10$~kG the probabilities of these atomic transitions remain unchanged.
Also, the frequency shifts of the transitions labeled $19$ and $20$ are simply linear versus magnetic $B$-field.
Such type of remarkable atomic transitions is absent for $D_1$ line in the case of  circular  polarized laser radiation.\\
iii) In order to determine theoretically the frequency positions of atomic transitions in the case
of $D_1$ line ($J=1/2$) the well-known Rabi-Breit formulas can be implemented~\cite{Var}, while they are not useful for $D_2$ line ($J=3/2$).

\section{Conclusion }
\indent We present the results of experimental and theoretical studies of $^{87}$Rb and $^{85}$Rb
$D_2$
line transitions in a wide range of magnetic fields 3~kG $ <B <7$~kG. It is experimentally demonstrated that in the case of $B >3$~kG and $\sigma^+$ polarized laser radiation, there remain only $20$ Zeeman transitions
while there are $60$ allowed Zeeman transitions at low $B$-field. In the case
of $B >4.5$~kG in the absorption spectrum these $20$ atomic transitions are regrouped to form two completely separate groups of $10$ atomic transitions each with the frequency slopes $s_1$ and $s_2$ correspondingly.
The frequency separation between the two groups increases with the magnetic field. The above mentioned peculiarities are the manifestation of hyperfine Paschen-Back regime. The implemented theoretical model very well describes the experiment.\\
\indent Possible applications of the $\lambda/2$-method can be mentioned: (i) making a reference at a strongly (up to $\pm15$~GHz)
shifted frequency with respect to the initial atomic levels of Rb with the use of a permanent magnet as well as
frequency locking of diode-laser radiation to these resonance~\cite{Sarg2012}; (ii) designing a magnetometer for measuring
strongly inhomogeneous magnetic fields with a high spatial resolution~\cite{Sargsyan1}; (iii) designing an optical insulator
based on Rb vapor with the use of the Faraday effect in strong magnetic fields ($\sim6$~kG), as was implemented in~\cite{Weller1}.\\
\indent Taking into account that the probability of atomic transitions increases with $B$, we do not expect any
limitations for using the $\lambda/2$-method at $B > 10$~kG.
\section{Acknowledgement}
\indent The research leading to these results has received funding from the European Union FP7/2007-2013
under grant agreement N$^{\circ}$ $295025$-IPERA. The research was conducted in the scope of the International
Associated Laboratory IRMAS (CNRS-France \& SCS-Armenia).
A.S., G.H., and D.S. acknowledge the support of the State Committee for Science, Ministry of
Education and Science of the Republic of Armenia (project no. SCS 13-1C029).


\end{document}